\documentstyle[aps]{revtex}

\newcommand{\bra}[1]{\left\langle #1 \right|}
\newcommand{\ket}[1]{\left| #1 \right\rangle}

\begin{document}
\draft
\title{Width of the $\Delta$ Resonance in Nuclei}
\author{M. Hjorth-Jensen}
\address{Fysisk Institutt, Universitetet i Oslo, N-0316 Oslo, Norway}
\author{H.\ M\"{u}ther}
\address{Institut f\"{u}r Theoretische Physik, Universit\"{a}t T\"{u}bingen
D-72076 T\"{u}bingen, Germany}
\author{A.\ Polls}
\address{Departament d'Estructura i Constituentes de la Materia,
Universitat de Barcelona, E-08028 Barcelona, Spain}

\maketitle


\begin{abstract}
In this work we evaluate the imaginary
part of the isobar $\Delta$ self-energy  $\Sigma_{\Delta}$
from the two-body absorption process $\Delta+N\rightarrow 2N$.
This contribution is calculated using a recently developed
non-relativistic scheme, which allows for  an evaluation of the
self-energy with a basis of single-particle states appropriate
for both bound hole states  and for particle states in the continuum.
In order to test the medium dependence of the self-energy, we calculate
the two-body absorption term $\Sigma_{\Delta}^{A2}$ for several finite
nuclei with $N=Z$, i.e.\ $^{16}$O, $^{40}$Ca and $^{100}$Sn.
The resulting self-energy, which is energy dependent and
non-local, is compared with a simple parameterization derived from
nuclear matter.
\end{abstract}

\pacs{14.20.Gk, 21.60.-n,24.10.Cn}

The self-energy of the $\Delta$ resonance is central
in the understanding of nucleus-nucleus collisions,
pion-nuclear processes and photonuclear
reactions at intermediate energies, see e.g.\ Ref.\ \cite{gaarde91}
for a recent review.
The self-energy $\Sigma_{\Delta}$ is given by a real part and an imaginary part
\begin{equation}
  \Sigma_{\Delta}(\omega ,k,k')=
  Re\Sigma_{\Delta}(\omega ,k,k')
  +iIm\Sigma_{\Delta}(\omega ,k,k'),
\end{equation}
where $\omega$ and $k,k'$ are the energy and momenta
of the isobar, respectively.

In a microscopic derivation of the self-energy, it is common to represent
various contributions to the self-energy by way of Feynman diagrams.
Typical examples are shown in Fig.\ \ref{fig:delself}.
Experimental data for pion scattering and absorption
at various energies covering
the isobar resonance region, suggest that
the dominant process for absorption of pions at low energies
is represented by  diagram (a), which couples the isobar to two-nucleon
one-hole states ($\Delta + N \rightarrow 2N$).
Within the terminology of the isobar-hole model \cite{ow79,ew88,otw83},
diagram (b) is then supposed to represent the rescattering of e.g.\ a real
pion, so-called reflection contributions to quasi-free scattering.
Diagram (c) is an example of a
self-energy contribution arising from three-body
absorption mechanisms.

Unfortunately, there
is no such thing as an experimental measurement of the
$\Delta$ self-energy, although
indirect information about the self-energy
can be derived from e.g.\ pion-nucleus scattering. In the extensive
analyses of Ref.\ \cite{hirata80}, contributions arising from diagrams
like (a) and (b) in Fig.\ \ref{fig:delself}, are represented by way
of a $\Delta$-spreading potential, fitted to provide best results for
pion-nucleus elastic scattering.

In Ref.\ \cite{os87}, a parameterization for the imaginary part of the
$\Delta$ self-energy is obtained in nuclear matter by considering the
contributions from the diagrams  in Fig.\ \ref{fig:delself}, accounting
thus for quasielastic corrections, two-body and three-body
absorption. The nuclear matter results are then compared to the
$\Delta$-spreading potential from the empirical determination in
Ref.\ \cite{hirata80} by allowing for a density dependent self-energy,
given by the analytical expression
\begin{equation}
Im\Sigma_{\Delta} =
Im\Sigma_{\Delta}^{A2} +Im\Sigma_{\Delta}^{A3} +Im\Sigma_{\Delta}^{Q},
\label{eq:oset}
\end{equation}
The term ${Q}$ accounts for the quasielastic part
while ${A2}$ is the
two-body absorption part.
${A3}$ represents three-body absorption.
A numerical parameterization for these terms was given by Oset and Salcedo
\cite{os87}, while analytical expressions based on the results of Ref.\
\cite{os87} were recently presented by Nieves {\em et al.} \cite{nieves93}.
In the latter work the $\Sigma_{\Delta}^{A2}$ term of Eq.\ (\ref{eq:oset})
has been calculated for a ``typical'' $\Delta$, which is excited in the
$\Delta$-hole model, when a pion of a certain kinetic energy is
absorbed. This parameterization reads
\begin{equation}
Im\Sigma_{\Delta}^{A2}(x)=Im \alpha_{\Delta}(x)\frac{1}{\beta (x)}
arctg\left(\beta (x)(\rho/\rho_0)\right),\label{eq:imdelt}
\end{equation}
with $x=\frac{T_{\pi}}{m_{\pi}}$, $T_{\pi}$ and $m_{\pi}$ being the kinetic
energy and
mass of the pion, respectively. $\rho$ is the density of particles and
$\rho_0$
the corresponding density at nuclear matter saturation.
Further, the function
$Im\alpha_{\Delta}$  is
\begin{equation}
Im \alpha_{\Delta}(x)=-38.3\left(1-0.85x+0.54x^2\right) \mbox{MeV} ,
\end{equation}
and
\begin{equation}
\beta(x)=2.72-4.07x+3.07x^2.
\end{equation}
The parameterization of eq.(\ref{eq:imdelt}) can be interpreted as the
width of a $\Delta$, which is excited when an energy $\omega =T_{\pi}
+ m_{\pi}$ is deposited at a nucleon in nuclear matter of density
$\rho$.

The intention behind this brief report is to study the $\Sigma_{\Delta}^{A2}$
term directly for finite nuclei,
since this contribution is expected to be the dominant one at energies
below the isobar resonance \cite{os87}. For that purpose  we will here
employ a recently developed method to calculate
the self-energy of the $\Delta$ in a finite nucleus.
Microscopic calculations of
e.g.\ the nucleon or $\Delta$ self-energy have preferentially been carried
out in nuclear matter, the
results of Ref.\ \cite{os87} being one example. One of the advantages of
studies in nuclear matter is the possibility to describe the single-particle
wave functions by plane waves. For a microscopic calculation in finite
nuclei one has to take into account the fact that one needs different
representations for bound hole  states and particle states in the
continuum. A method which allows for this has recently
been developed \cite{bm89,hbmp93}.
The bound hole states are described in terms of harmonic oscillator
(h.o.) wave functions while particle states are given by plane waves.
The basic ingredients in our microscopic calculation of
$\Sigma_{\Delta}^{A2}$ are briefly outlined below.
The finite nuclei we consider are $^{16}$O, $^{40}$Ca and $^{100}$Sn.

In order to calculate $\Sigma_{\Delta}^{A2}$, we need first to define
the transition potential $V_{NNN\Delta}$
for the $\Delta +N\rightarrow 2N$ process.
For a nucleon and an isobar $\Delta$
interacting through the exchange of $\pi$ plus $\rho$ mesons,
the transition potential $V_{NNN\Delta}$
is usually written, in the static
 non-relativistic
limit, as
\cite{ew88}
\begin{equation}
V_{NNN\Delta}({\bf k})={\displaystyle -\left\{
D_{\pi}^{N\Delta}({\bf k})\frac{f_{\pi NN}f_{\pi
 N\Delta}}{m_{\pi}^{2}}
\mbox{\boldmath $\sigma$} \cdot {\bf k}{\bf S}\cdot
{\bf k}
+D_{\rho}^{N\Delta}({\bf k})\frac{f_{\rho NN}f_{\rho N\Delta}}{m_{\rho}^{2}}
\mbox{\boldmath $\sigma$}\times {\bf k}\cdot {\bf S}\times {\bf k}
\right\}
\mbox{\boldmath $\tau$}{\bf T},}   \label{eq:vndel}
\end{equation}
with $S(T)$ the transition matrix which creates a spin (isospin)
$3/2$ object from a spin (isospin) $1/2$ one.
The meson propagator $D_{\pi , \rho}({\bf k})$ is defined as in Ref.\
\cite{hbmp93} as
\begin{equation}
D_{\pi , \rho}^{N\Delta}({\bf k})=
\frac{1}{2}\left(\frac{1}{m_{\pi ,\rho}^{2}+{\bf k} ^{2} }+
\frac{1}{m_{\pi , \rho}^{2}+{\bf k} ^{2} +
m_{\pi , \rho}(m_{\Delta}-m_{N})}\right).
\end{equation}
The coupling constant for the $\pi$
meson is given by a relation obtained from
the non-relativistic quark model \cite{bw75}
\begin{equation}
f_{\pi N\Delta} = \frac{6}{5}\sqrt{2}f_{\pi NN}=
\frac{6}{5}\sqrt{2}g_{\pi NN}\frac{m_{\pi}}{2m_{N}},
\end{equation}
and similarly for the $\rho$ meson we have
\begin{equation}
f_{\rho N\Delta} =\frac{f_{\pi N\Delta}}{f_{\pi NN}}
g_{\rho NN}\frac{m_{\rho}}{4m_{N}}\left(
1+\frac{f_{\rho NN}}{g_{\rho NN}}\right),
\end{equation}
with
\begin{equation}
f_{\rho NN} = \sqrt{4\pi}g_{\rho NN}\frac{m_{\rho}}{m_{N}}\left(
1+\frac{f_{\rho NN}}{g_{\rho NN}}\right).
\end{equation}

In addition we include monopole form factors in order to regularize the
potentials at short distances.
The cutoff masses are $\Lambda_{\pi}=1.2$ GeV and $\Lambda_{\rho}=1.3$
GeV, while the coupling constants are $g_{\pi NN}^2/4\pi=14.6$ and
$g_{\rho NN}^2/4\pi=0.95$, which are equal to the parameters which define
the Bonn B nucleon-nucleon potential $V_{NN}$ of table A.2 in Ref.\
\cite{mac89}. Further, $\frac{f_{\rho NN}}{g_{\rho NN}} =6.1$.
The Bonn B potential is used to calculate the
$G_{N\Delta}$-matrix, the next ingredient in our calculations.
To calculate the  $G_{N\Delta}$-matrix, we need first to evaluate the
nucleon-nucleon $G$-matrix $G_{NN}$. This is done by solving the
Bethe-Goldstone equation
\begin{equation}
  G_{NN}(\Omega)
  = V_{NN}+V_{NN}
  Q\frac{1}{\Omega - QH_0Q}Q G_{NN}(\Omega).
  \label{eq:gmat}
\end{equation}
Here $V_{NN}$ is the free nucleon-nucleon interaction. In this work $V_{NN}$ is
defined by the parameters of the Bonn B potential in table A.2 of Ref.\
\cite{mac89}. The term $H_0$ is the unperturbed hamiltonian.
This equation is solved with an angle-average
nuclear matter Pauli operator $Q$
with a fixed starting energy $\Omega=-10$ MeV and a Fermi momentum
$k_F=1.4$ fm$^{-1}$. From the nucleon-nucleon $G_{NN}$ matrix, we
can evaluate the $G_{N\Delta}$ matrix \cite{hbmp93} through the relation
\begin{equation}
  G_{N\Delta}(\Omega)
  = V_{NNN\Delta}+V_{NNN\Delta}
  Q\frac{1}{\Omega - QH_0Q}Q G_{NN}(\Omega).
  \label{eq:gndel}
\end{equation}
Having accounted for the short-range correlations through the introduction
of the $G_{N\Delta}$-matrix, we are then able to set up the expression
for the imaginary part of $\Sigma_{\Delta}^{A2}$
\begin{equation}
  \begin{array}{ll}
  Im\Sigma_{\Delta}^{A2}(j_bl_bk_{b}k_{a}\omega) =
  &{\displaystyle -\frac{1}
  {2(2j_b+1)}\sum_{n_{h}l_{h}j_{h}}
  \sum_{JT}\sum_{lLS{\cal J}}\int k^{2}dk\int K^{2}dK\hat{J}\hat{T}}\\&\\
  &\times \bra{k_{a}l_{b}j_{b}n_{h}l_{h}j_{h}JT}
  G_{N\Delta}\ket{klKL({\cal J})SJT}\\&\\
   &\times\bra{klKL({\cal J})SJT}
   G_{N\Delta}\ket{k_bl_bj_bn_{h}l_{h}j_{h}JT}\\&\\
   &\times\pi\delta(\omega + \varepsilon_{h}-\frac{K^2}{4M_N}
    -\frac{k^2}{M_N}),
\end{array} \label{eq:2p1hbb}
\end{equation}
The single-hole energy $\varepsilon_{h}$ is given by the eigenvalues
of the harmonic oscillator minus a constant shift to place the Fermi
energy at zero, while the energies of the particle states are
represented by the pure kinetic energy.
The variables $k, K$ are the relative
and center-of-mass momenta of the intermediate particle states $p_1$ and
$p_2$ in Fig.\ 1. Further, $l$ and $L$ are the corresponding orbital
momenta of the relative and center-of-mass motion. $S$, $J$ and $T$ are the
total spin, total angular momentum and isospin, respectively. Finally,
$M_N$ is the average proton and neutron masses.
A h.o. single-particle state is defined by the
quantum numbers $n_hl_hj_h$, while
plane waves are defined by $k_al_aj_a$.
For further
details, see Ref.\ \cite{hbmp93}.
The energy variable $\omega$
refers to the energy of the $\Delta$ relative to the mass of a nucleon.
Only positive energies $\omega$ contribute, as can be deduced from the
$\delta$ function in Eq.\ (\ref{eq:2p1hbb}).

To study the medium dependence of $Im\Sigma_{\Delta}^{A2}$, we evaluate
Eq.\ (\ref{eq:2p1hbb}) for the nuclei $^{16}$O, $^{40}$Ca and $^{100}$Sn.
The medium dependence of Eq.\ (\ref{eq:2p1hbb}) is accounted for
by the summation over single-hole states, represented by the
$0s_{1/2}$, $0p_{1/2}$ and $0p_{3/2}$ single-hole states
in $^{16}$O,
$0s_{1/2}$, $0p_{1/2}$, $0p_{3/2}$
$1s_{1/2}$, $0d_{3/2}$ and $0d_{5/2}$ single-hole states
in $^{40}$Ca and
$0s_{1/2}$, $0p_{1/2}$, $0p_{3/2}$
$1s_{1/2}$, $0d_{3/2}$, $0d_{5/2}$
$1p_{1/2}$, $1p_{3/2}$,
$0f_{5/2}$, $0f_{7/2}$ and $0g_{9/2}$
single-hole states
in $^{100}$Sn. Moreover, the oscillator parameters used in the
calculation of the single-hole wave functions are $1.72$ fm for
$^{16}$O, $2.04$ fm for
$^{40}$Ca  and $2.20$ fm for
$^{100}$Sn.

In the discussion presented here, we only consider $\Delta$ isobar
states with orbital angular momentum $l_{b}=0$. A Fourier transformation of
$Im\Sigma_{\Delta}^{A2}$ in Eq.\ (\ref{eq:2p1hbb}) leads to an
imaginary part, which depends on energy $\omega$ and is non-local in the
coordinate $r$, the distance from the center of the nucleus. From the
inspection of this function we observe that the non-locality is weak
in the sense that it is different from zero only for distances $r$ and
$r'$, which are close to each other. Therefore it makes sense to look
at the local component of $Im\Sigma_{\Delta}^{A2}$ for the various
energies as a function of the distance $r$ \cite{bbmp92}. As an example
we present in the left part of Fig.\ 2 this local approximation
obtained for $^{40}$Ca. The shape of these functions is not really
identical to a Woods-Saxon shape or a conventional density distribution.
In particular at lower energies
($\omega$ below 200 MeV) one observes a clear surface contribution to
$Im\Sigma_{\Delta}^{A2}$. Similar results are also obtained for the
other nuclei (see right side of Fig.\ 2).

Finally, in Fig.\ 3, we compare the imaginary part of the $\Delta$
self-energy calculated in the local approximation for a typical radius
of $r$=1.5 fm at various energies $\omega$
with the parameterization of Nieves et al.~\cite{nieves93}. This
comparison must be considered with some care. As discussed above, the
parameterization of eq.(\ref{eq:imdelt}) represents the average imaginary part
of a $\Delta$, which is typically excited, when a pion is absorbed in
nuclear matter of density $\rho$, depositing an energy $\omega = T_{\pi}
+ m_{\pi}$. On the other hand, the results for finite nuclei show a
non-trivial radial dependence and Fig.\ 3, just displays results for
one ``typical'' radius. The nuclear matter parameterization is presented
for a density $\rho$= 0.75 $\rho_{0}$ which is the average density of
nucleons in $^{100}$Sn. The agreement between the microscopic
calculation for this nucleus and the parameterization,
which is based on studies of nuclear matter is remarkable. This is
true for both the absolute value as well as the shape of the
energy-dependence. Only for the lightest nucleus which we considered,
$^{16}$O, the calculated width lies considerably below the
parameterization.

In conclusion we would like to point out that our microscopic
evaluation of the $\Delta$-spreading potential in finite nuclei
supports the parameterization of \cite{nieves93}, which is based on
studies of nuclear matter. For low energies, however, it may be
important to consider a surface enhancement of the imaginary part in
the self-energy of the $\Delta$ isobar.

\bigskip
This work has been supported by the NorFA (Nordic Academy for Advanced
Study)
through grant $\#93.40.018/00$ , the ``Bundesministerium f\"ur
Forschung und Technologie'' (06 T\"u 736, Germany) and by the contract DGICYT
no.~PB92-0761 (Spain).


\clearpage
\begin{figure}
\caption{Example of diagrams which arise in the evaluation of the
self-energy of the isobar. (a) is the two-body absorption term evaluated
in this work, (b)
is an example of a so-called reflection contribution to quasi-free
scattering while (c) stands for a three-body contribution. The wavy line
represents the $G_{N\Delta}$-matrix, the single line is a nucleon while
the double stands for the isobar $\Delta$.}
\label{fig:delself}
\end{figure}

\begin{figure}
\caption{Local representation of the imaginary part of
$-\Sigma_{\Delta}^{A2}$ as function of the distance $r$ from the center
of the nucleus. In the left part of the figure results are shown for
the nucleus  $^{40}$Ca, considering the energies $\omega$ = 100, 200
and 400 MeV. In the right part of the figure the results are displayed
for the nuclei $^{16}$O, $^{40}$Ca and $^{100}$Sn assuming an energy
$\omega$=300 MeV.}
\label{fig:selfa2}
\end{figure}

\begin{figure}
\caption{Strength of the imaginary part of $-\Sigma_{\Delta}^{A2}$ at
various energies. Results obtained for the local representation at
$r$=1.5 fm, derived from the microscopic calculation of the $\Delta$
self-energy in the finite nuclei $^{16}$O, $^{40}$Ca and $^{100}$Sn are
compared to the parameterization of Nieves et al. [7], displayed in
eq.(3) for a density $\rho=0.75\rho_{0}$. This parameterization has
been extrapolated from $\omega=m_{\pi}$ to $\omega$=0.}
\label{fig:selfws}
\end{figure}

\end{document}